\documentclass{Interspeech}

\interspeechcameraready 

\title{Domain Adaptation Method and Modality Gap Impact in Audio-Text Models for Prototypical Sound Classification}

\author[affiliation={1}]{Emiliano}{Acevedo}
\author[affiliation={1,2}]{Martín}{Rocamora}
\author[affiliation={3}]{Magdalena}{Fuentes}

\affiliation{Facultad de Ingeniería}{Universidad de la República}{Uruguay}
\affiliation{Music Technology Group}{Universitat Pompeu Fabra}{Spain}
\affiliation{MARL-IDM}{New York University}{USA}
\email{eacevedo@fing.edu.uy, martin.rocamora@upf.edu, mfuentes@nyu.edu}

\usepackage{comment}
\usepackage{cite}
\usepackage{hyperref}
\usepackage{url}

\begin{document}

\maketitle

\begin{abstract}
    
    Audio-text models are widely used in zero-shot environmental sound classification as they alleviate the need for annotated data. However, we show that their performance severely drops in the presence of background sound sources. Our analysis reveals that this degradation is primarily driven by SNR levels of background soundscapes, and independent of background type. To address this, we propose a novel method that quantifies and integrates the contribution of background sources into the classification process, improving performance without requiring model retraining. Our domain adaptation technique enhances accuracy across various backgrounds and SNR conditions. Moreover, we analyze the modality gap between audio and text embeddings, showing that narrowing this gap improves classification performance. The method generalizes effectively across state-of-the-art prototypical approaches, showcasing its scalability and robustness for diverse environments.

\end{abstract}

\section{Introduction}
\label{sec:intro}

Audio-text models (ATMs), i.e. embedding models with semantic knowledge of language and sound, have become popular in zero-shot (ZS) sound classification as they relieve the need for collecting and annotating audio recordings \cite{guzhov2022audioclip, elizalde2023clap}. 
Aligning audio and text modalities facilitates prototype approaches, where \textit{anchors}, specific embedding points of one modality, act as class prototypes for the other within a shared latent space. 
A common method uses text \textit{anchors} to define audio class prototypes and then classify unlabeled audio based on its proximity to the nearest prototype.
Such models have potential in applications such as noise pollution analysis \cite{rey2016analysis}, surveillance systems \cite{7321013}, hearing aid technology \cite{10.1145/3597638.3608431}, and smart city initiatives \cite{RENAUD2023119568}. Among audio-text models, LION-CLAP \cite{laionclap2023} stands out in multiple tasks \cite{soundsimilarity2024} as one of the best-performing due to training on one of the largest available audio-text datasets, and careful selection of hyperparameters for both audio and text encoders.

Despite the real-world nature of sound classification (i.e. sounds are rarely encountered in isolation but within a background soundscape), most ATMs have been used for ZS classification with isolated sound sources. In this work, we show that the presence of background sound sources significantly degrades ATMs performance. Recently, domain adaptation (DA) techniques for ATMs have been proposed \cite{deshmukh2024domain, hanif2024palm}, focusing only on optimizing the text encoder's input or feature space by learning domain-specific tokens. 
Furthermore, as noted in \cite{wu2023audio}, ATMs struggle to fully understand natural language, particularly contextual concepts such as sequential or concurrent sound events. Our analysis further reveals that treating the background as a high-level text concept (e.g., ``the sound of a park") is misleading and often results in misclassifications driven by background sound sources. These findings underscore the need for more precise characterizations of background soundscapes to achieve effective DA in these scenarios.
Motivated by these limitations, we propose a method that quantifies the contribution of background sound sources in an audio recording to improve ZS classification. By treating the background as a combination of individual sounds, our method consistently improves performance across various types and amplitude levels of background soundscapes without requiring model retraining.

Besides the challenges of real-world sound classification, audio-text models suffer from a common issue of multimodal models, the modality gap (i.e. a semantic misalignment of modalities in the embedding space). This is a known consequence of models trained with contrastive loss \cite{fahim2024its, NEURIPS2022_702f4db7}. 
Previous solutions, such as projection-based and nearest-neighbor decoding \cite{kouzelis2023weakly}, have addressed this issue. Building on these, we show that the recent method proposed by \cite{kushwaha2023multimodal}, which leverages unlabeled audio, is similarly effective for narrowing the gap, resulting in improved performance.

This work makes three key contributions: (i) a quantification of background sound sources' influence in zero-shot classification, (ii) a novel method that incorporates background profiles to improve performance in domain shifts, and (iii) an analysis of the modality gap's impact on sound classification.  
Our code is made available as open source for further research.
\footnote{\href{https://github.com/eacevedo1/AudioText-ContextDA.git}{\texttt{github.com/eacevedo1/AudioText-ContextDA.git}}}

\section{Impact of Background Sounds}
\label{sec:data_augmentation}


To assess the impact of background sound on ZS sound classification using ATMs, we synthetically generated soundscapes using the \texttt{Scraper} library \cite{salamon2017scaper}. The foreground samples were drawn from UrbanSound8k \cite{salamon2014dataset}, while the background soundscape samples were taken from TAU Urban Acoustic Scenes 2019 \cite{Mesaros2018_DCASE}. 
We generated new soundscapes by pairing each UrbanSound8k foreground sound with one random background sample from three indoor (Shopping Mall, Metro Station, Airport) and three outdoor (Park, Public Square, Street Traffic) environments, varying the signal-to-noise ratio (SNR) across three levels, namely 6, 8, and 10 dB. The selected values ensure that foreground sounds remain salient sounds, while preserving the background's texture-like qualities. Values below 6 dB make the foreground indistinguishable from the background, whereas values above 10 dB minimize the impact of the background, consistent with the insights from previous work \cite{radford2023robust}.

\begin{figure} [ht]
    \centering
    \includegraphics[width=1\linewidth]{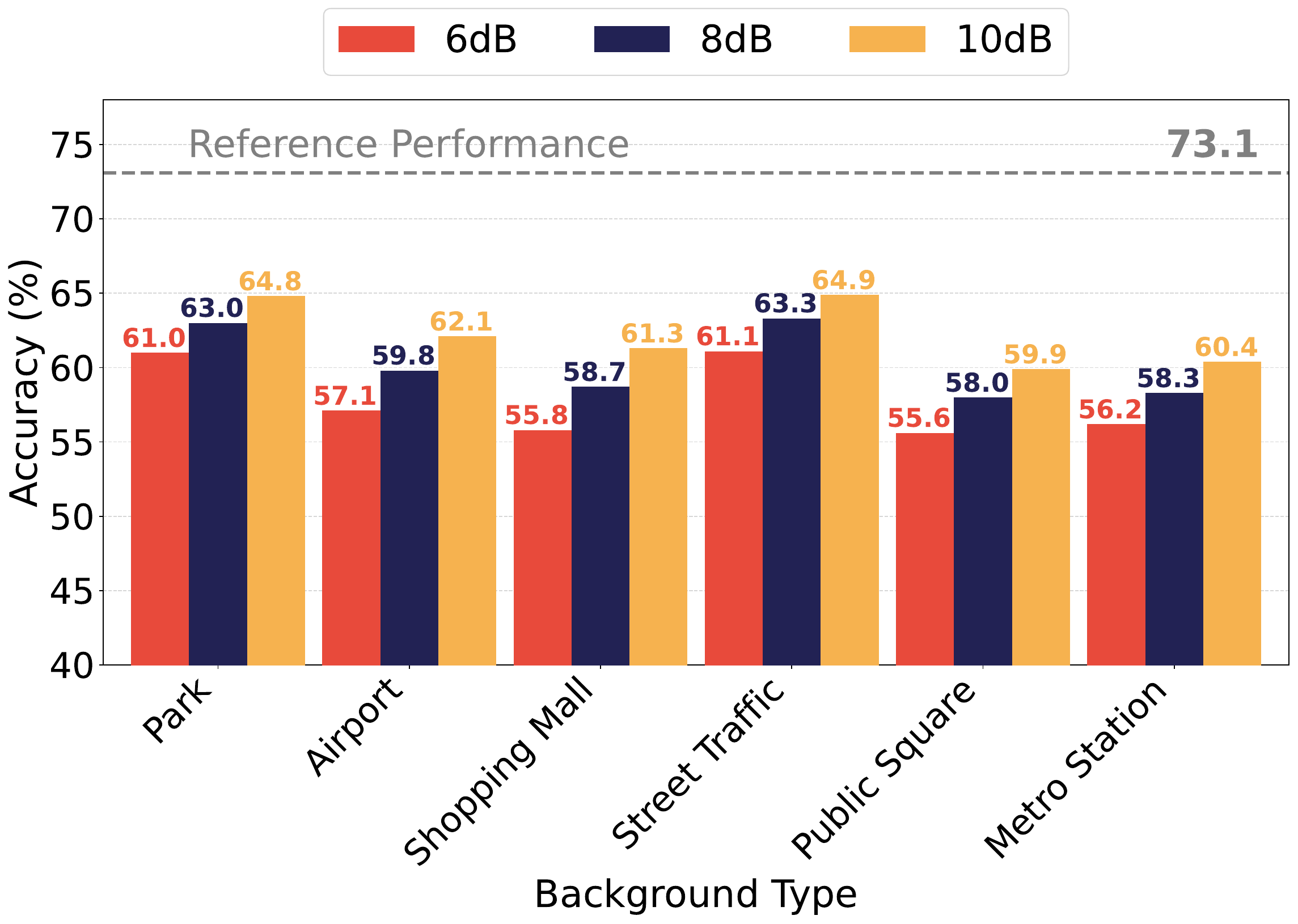}
    \caption{Impact of background soundscape on zero-shot sound classification accuracy across different acoustic environments. The plot compares model performance under 6 dB, 8 dB, and 10 dB SNR conditions against the reference performance (i.e. isolated sources).}
    \label{fig:impact-of-background}
\end{figure}

Figure \ref{fig:impact-of-background} shows the classification performance of different sources from UrbanSound8k in the presence of background, compared to a reference performance of the sounds in isolation.
In every case, we observe a notable decline in performance once the background soundscape is introduced.
Performance degradation is tied to SNR levels—lower SNR (higher background presence) results in poorer performance. This decrease is consistent across both indoor and outdoor environments, with SNR being the primary factor affecting the model.


\begin{figure*} [ht] 
  \includegraphics[width=\textwidth]{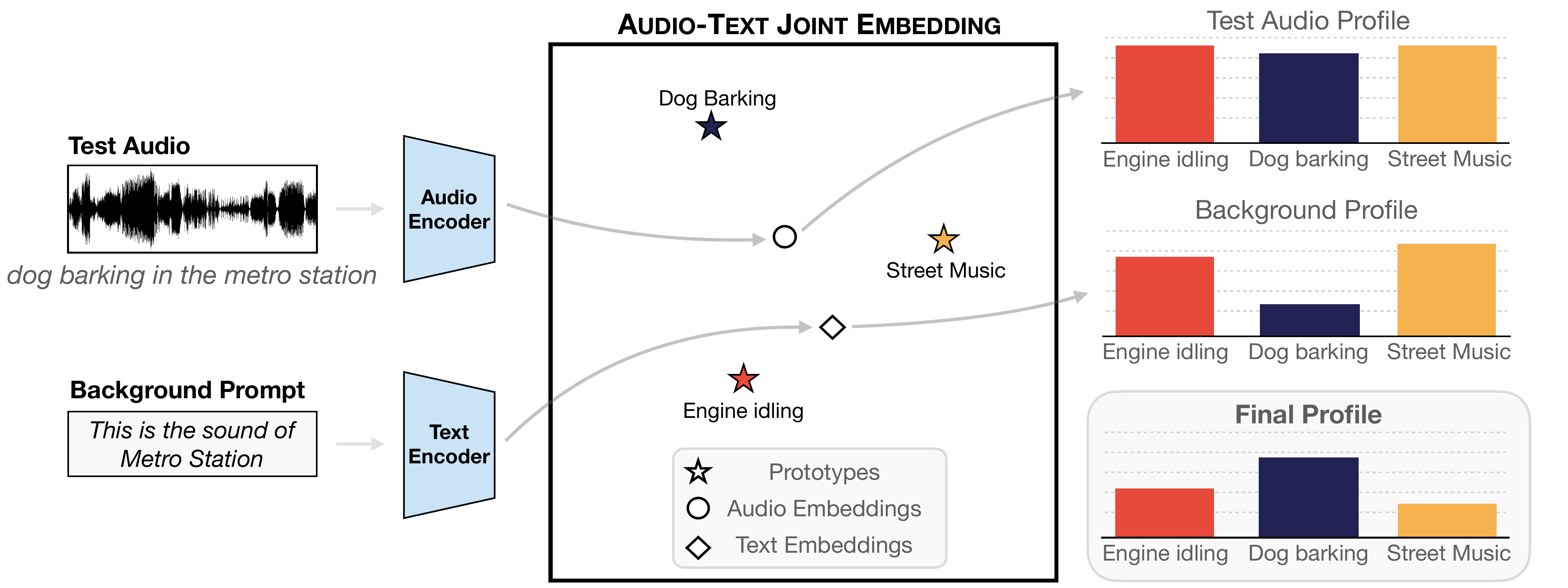}
  \caption{ Proposed method for domain adaptation in prototypical sound classification. The method quantifies the background's contribution to the test audio soundscape and refines class estimation by subtracting its influence. First, the test audio profile is calculated by measuring the similarity between the test audio embedding and the class prototypes. Next, a background profile is generated, either from text descriptions or audio recordings, and subtracted from the test audio profile to minimize the impact of background classes. 
  }
  \label{fig:method-proposed}
\end{figure*}

In addition to evaluating downstream task performance, we measured the cosine distance between text \textit{anchors} and generated soundscapes, which averaged 0.76--slightly higher than the 0.73 observed with isolated foreground sounds. To investigate whether explicitly providing background information could improve performance, we modified the prompt to include the background type (e.g., \texttt{'This is a sound of LABEL in the BACKGROUND TYPE'}). This adjustment improved the alignment between text \textit{anchors} and audio embeddings, reducing the cosine distance to 0.67. However, this improvement did not result in better classification performance across all background types.
Moreover, we conducted ZS classification on the background soundscapes alone, achieving an accuracy of only 33.0\%, which is significantly lower than 63.3\% reported in the acoustic scene classification baseline \cite{Mesaros2018_DCASE}. This suggests that the model struggles to differentiate between various background types. These results imply that the concept of ``background" is not well represented as a high-level concept within the model. 

However, we observed that some background information was encoded within the model. 
For instance, when queried with \emph{Airport} soundscapes, the model returned classes such as 
\emph{bell}, \emph{pedestrians} and \emph{phone ringing}. Similarly, the prompt \texttt{'Public square sound in the background'} yielded results like \emph{people talking}, \emph{pedestrians}, and \emph{footsteps}. These findings suggest that the model encodes some relationships between background soundscapes and individual classes.
In addition, we observed that the model's classification errors were systematically linked to the background sound sources. For example, when adding \emph{Street Traffic} soundscapes, the model top confusions are with \emph{sirens}, \emph{engine idling}, and \emph{car horns}, or when adding \emph{Shopping Mall} the model main confusions are \emph{street music}, \emph{air conditioner} and \emph{children playing}. Adding a background soundscape transforms the audio into polyphonic, which poses significant challenges for prototypical approaches, as they struggle to represent overlapping sources as a single point in the latent space. This limitation arises from the uniform aggregation of both temporal and timbral information, which fails to capture the inherent complexity of polyphonic acoustic scenes, as previously noted in \cite{dogan2024multi}. 
Each individual background sound source influences the position of the query audio embedding with respect to different text \textit{anchors} (i.e. the classes),
diminishing confidence in a single foreground sound. This leads to unintended confusion between classes in the task and those present in the background soundscape that are acoustically similar or overlapping.  

To study the effect of polyphony, we generated a dataset of 1000 samples without background, varying the number of classes per audio (C/A) by combining UrbanSound8k foreground sounds. We performed multi-label sound classification on this dataset, and results are shown in Table \ref{tab:multi-label-experiment}. We observe that with a fixed classification threshold, the average number of predicted classes per audio (Pred. C/A) decreases as the complexity of the audio scene increases, resulting in lower accuracy. We include both ZS and supervised approaches (i.e., using the class centroid based on annotations) to demonstrate that this effect is not specific to the audio-text model but is inherent to the prototype-based methods.

\begin{table}[ht]
\fontsize{9pt}{9pt}\selectfont
\centering
\begin{tabular}{ccccc}
\toprule
\textbf{} & \multicolumn{2}{c}{\textbf{Zero-Shot}} & \multicolumn{2}{c}{\textbf{Supervised}} \\
\cmidrule(lr){2-3} \cmidrule(lr){4-5}
\textbf{C/A} & \textbf{Accuracy} & \textbf{Pred. C/A} & \textbf{Accuracy} & \textbf{Pred. C/A} \\
\midrule
1 & 59.5 & 0.9 & 71.4 & 0.8 \\
2 & 50.8 & 0.8 & 43.7 & 0.6 \\
3 & 48.5 & 0.6 & 39.3 & 0.4 \\
\bottomrule
\end{tabular}
\caption{Performance comparison between zero-shot (text \textit{anchors}) and Supervised (class centroid) modes on multi-label audio classification with varying numbers of classes per audio (C/A).}
\label{tab:multi-label-experiment}
\end{table}
\vspace{-0.5cm}


In summary, our experiments show that ATMs struggle to represent background as a coherent concept, making DA difficult. Adding background soundscapes turns audio into polyphonic, causing confusion and reducing classification confidence, which ultimately degrades model performance. Hence, to effectively address DA, the background should be treated as a combination of sound sources rather than a standalone concept, reducing its impact on the final classification.

\section{Method Proposed}
\label{sec:method}

Sound classification task in this context implies accurately identifying a foreground sound embedded within a background soundscape (e.g., detecting a dog barking in a metro station soundscape). The audio to be classified consists of a mixture of the foreground sound and the background soundscape, which we refer to as test audio soundscape as shown in Figure \ref{fig:method-proposed}. To address DA, we first define the profile concept as a vector of cosine similarity scores between a test audio embedding and the class prototypes. We begin the method by calculating the profile for the test audio soundscape.

Rather than just using this profile for classification, we refine it by subtracting the ``contribution'' of the background soundscape. To achieve this, we compute the background profile separately, using one of two approaches: (i) \textit{Text} approach, where textual prompts describing the background are compared to class prototypes to get the profile, and (ii) background \textit{Audio} approach, where an actual background audio recording is used to get the profile. Both approaches can use single or average multiple prompts or recordings for a more robust background profile. For instance, the best results for the \textit{Text} approach were obtained by averaging the profiles of three prompts: \texttt{'This is a sound of BACKGROUND TYPE'}, \texttt{'BACKGROUND TYPE sounds in the background'}, and \texttt{'This is a sound of BACKGROUND TYPE in the background'}.

After computing both the test audio and background profiles, we combine them to obtain a refined classification profile. This is done by subtracting the background profile from the soundscape profile, reducing the influence of background classes. The relationship is defined as:
\begin{equation}
\centering
    P_f = P_s - P_b \times \tau 
\end{equation}
where $P_s$ and $P_b $ represent the test audio profile and the background profile, respectively, and the parameter $\tau$ controls the degree of background influence on the final classification.

\begin{table*}[ht]
    \fontsize{10pt}{9pt}\selectfont
    \centering
    \begin{tabular}{ccccccc}
        \toprule
         & \multicolumn{3}{c}{\textbf{Zero-Shot}} & \multicolumn{3}{c}{\textbf{TGAP}} \\
        \cmidrule(lr){2-4} \cmidrule(lr){5-7}
        \textbf{Background} & Baseline & Text & Audio & Baseline & Text & Audio \\
        \midrule
        Park           & 61.0  & 62.2  & \textbf{64.3} & 65.7  & 67.7  & \textbf{72.2} \\
        Airport        & 57.1  & 58.8  & \textbf{60.3} & 65.1  & 66.3  & \textbf{70.2} \\
        Shopping Mall  & 55.8  & 57.5  & \textbf{60.2} & 63.2  & 65.2  & \textbf{68.7} \\
        Street Traffic & 61.1  & 63.5  & \textbf{66.8} & 68.8  & 70.6  & \textbf{72.6} \\
        Public Square  & 55.6  & 60.5  & \textbf{62.1} & 64.1  & 65.8  & \textbf{71.3} \\
        Metro Station  & 56.2  & 58.3  & \textbf{63.0} & 60.5  & 63.0  & \textbf{68.1} \\
        \cmidrule(lr){2-4} \cmidrule(lr){5-7}
        & \multicolumn{3}{c}{Zero-Shot Reference: 73.1} & \multicolumn{3}{c}{TGAP Reference: 76.9} \\
        \bottomrule
    \end{tabular}
    \caption{Performance comparison of \textit{Zero-Shot} and \textit{TGAP} approach across various context backgrounds using SNR of 6 dB. Using no domain adaptation (Baseline), domain adaptation methods using text (Text) and audio (Audio) to get the background profile.}
    \label{tab:results-table}
\end{table*}

\section{On the Audio-Text Modality Gap}
\label{sec:related_work}



So far, we have mentioned how prototypical approaches in ATMs struggle to represent polyphonic acoustic scenes and to effectively leverage text. Apart from this, ATMs also face the challenge of the \textit{modality gap},
wherein embeddings from different modalities are separated into distinct regions within the latent space. Despite its strong performance, LION-CLAP is not immune to the modality gap. On the UrbanSound8k dataset, we observed that the audio and text embeddings are linearly separable, with intra-modality distances consistently smaller than inter-modality distances. In particular, for any given class, the corresponding text \textit{anchors} are typically farther from the audio samples of that class than from audio centroids of other classes, highlighting the persistent nature of this issue.

\begin{figure}[ht]
    \centering
    \includegraphics[width=1\linewidth]{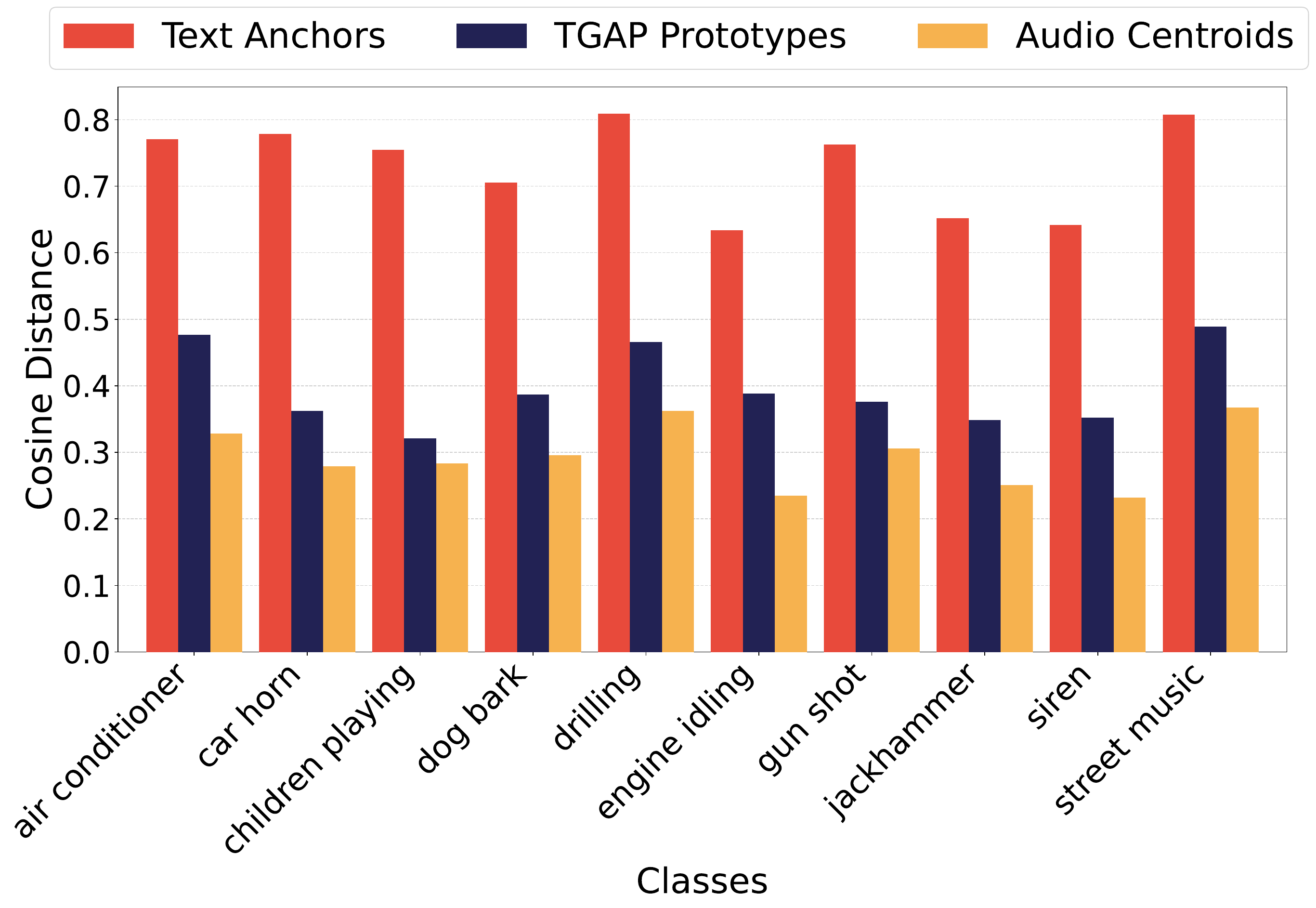}
    \caption{Comparison of cosine distances between audio samples and three types of class prototypes across multiple sound classes: text-based anchors (red), TGAP estimated centroids (blue), and real audio centroids (yellow) in UrbanSound8k.}
    \label{fig:anchors-distance}
\end{figure}

This modality misalignment suggests that relying solely on text \textit{anchors} might limit the performance of zero-shot sound classification. Motivated by this, 
we explored recent work \cite{kushwaha2023multimodal}, which introduced a multimodal prototypical approach that we called \textit{Text-Guided Audio Prototype} (TGAP).
This method generates class prototypes by retrieving the $N$ closest unlabeled audio embeddings to a text prompt and calculating their centroid, which becomes the prototype for classification. The method is fully unsupervised, without the need for re-training.
The TGAP method consistently improves sound classification performance over simple ZS classification across various datasets. To understand the source of this improvement, we examined the cosine distances from (i) text-based prototypes, (ii) TGAP-generated centroids and (iii) real centroids to corresponding class audio samples, as illustrated in Figure \ref{fig:anchors-distance}.

The experiment shows that text \textit{anchors} prototypes (red) have the highest cosine distances, indicating poor alignment with audio samples. TGAP prototypes (blue) significantly reduce this distance, demonstrating improved alignment with the audio embeddings. Real audio centroids (yellow), derived using labels, represent the case of no modality gap. Then, TGAP significantly reduces the modality gap compared to text-based prototypes. These findings highlight the critical role of modality alignment in zero-shot sound classification and the effectiveness of TGAP in narrowing the modality gap.

\section{Results and discussion}
\label{sec:results}

The results of our proposed method for ZS sound classification are shown in Table \ref{tab:results-table}. We compare three settings: no domain adaptation (Baseline), text-based adaptation (Text), and audio-based adaptation (Audio). Both adaptation approaches consistently improve performance across all background environments, with audio-based adaptation achieving the best results. 
The proposed method is flexible and effective with both text \textit{anchors} and TGAP prototypes. Applying the method to TGAP confirms its superior performance across different background types. Notably, the reduction in the modality gap provided by TGAP, also consistently boosts the performance compared to ZS. Audio-based adaptation also performs better with the TGAP method. We attribute this to the fact that actual background recordings offer a more precise and accurate representation than text descriptions. As mentioned earlier, text struggles to fully capture the background concept, making audio a more reliable source for adaptation.

\begin{figure} [ht]
    \centering
    \includegraphics[width=1\linewidth]{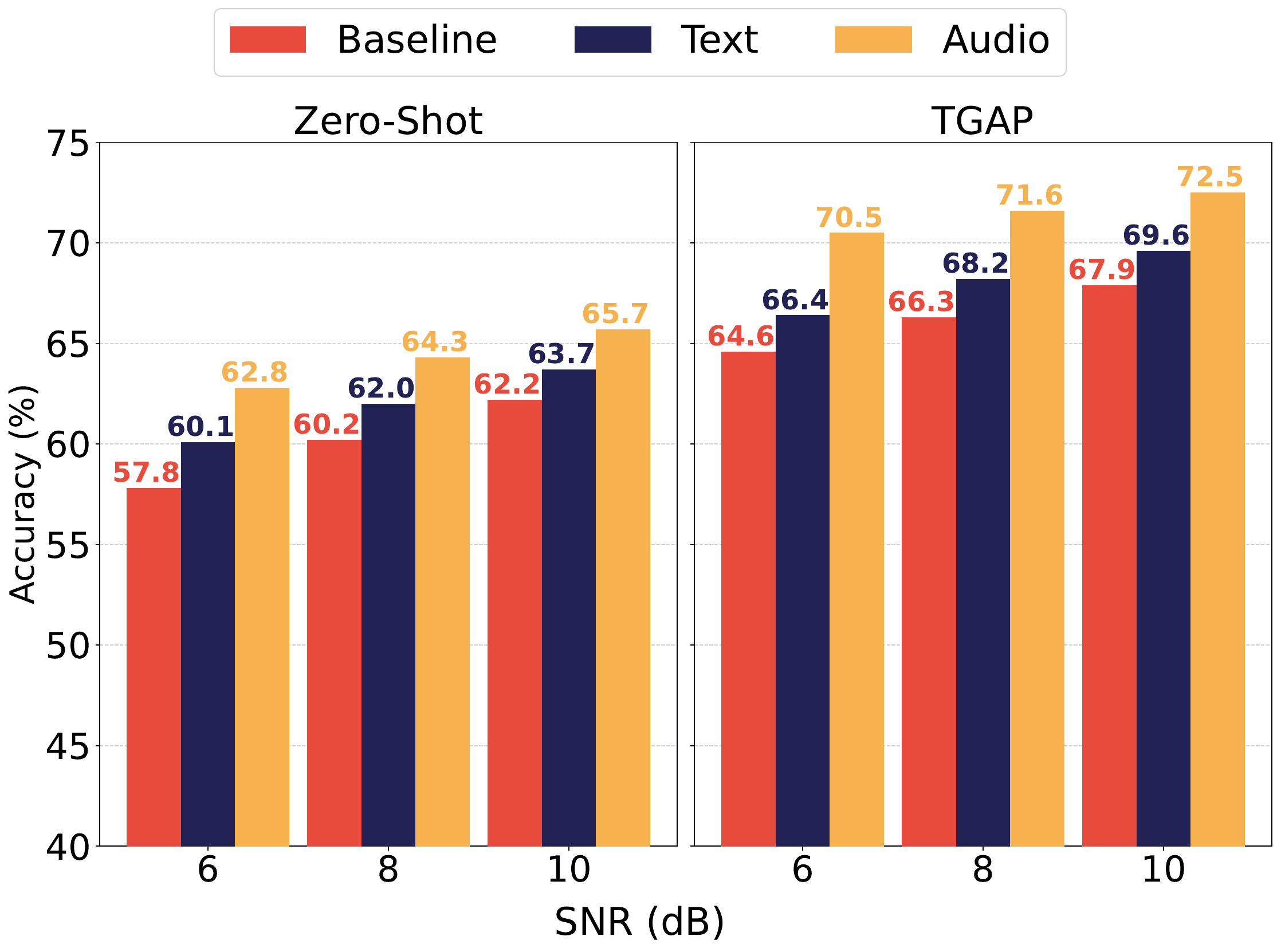}
    \caption{Average performance across all backgrounds under varying SNRs. The Zero-Shot method is shown on the left, and TGAP on the right, with Baseline, Text, and Audio evaluation setups in each.}
    \label{fig:domain-adaptation-snr}
\end{figure}

All the results above were computed with a fixed SNR of
6dB. Figure \ref{fig:domain-adaptation-snr} explores performance across varying SNRs (6, 8 and 10 dB) for both ZS and TGAP methods averaging over all background types. Our proposed adaptation consistently
improves performance at all SNR levels, with audio-based adaptation consistently yielding the highest gains. 
Regarding the parameter $\tau$, we optimized it using a grid search (0 to 1 in 0.1 steps), identifying 0.2 for text-based and 0.7 for audio-based adaptation as generally optimal. These values achieved a balance in background subtraction, as lower $\tau$ underestimated background effects, while higher $\tau$ overcompensated, reducing performance. The optimal value showed slight variations depending on the specific context and SNR's.



While the synthetic dataset allowed precise control over factors such as sound sources, polyphony, and SNR's, we also validated the generalization of our approach using the TUT datasets (DCASE SED Challenge 2016 \cite{Mesaros2016_EUSIPCO} and 2017 \cite{DCASE2017challenge}), which introduced real-world complexity with labeled foreground and background audio. The audio samples were segmented into 10-second chunks, with labels assigned based on annotations, and chunks without any class presence were categorized as background for the audio-based adaptation. Due to the dataset's structure, TGAP could not be applied, as isolated class instances within each chunk were unavailable. Applying the zero-shot method, the results obtained were 32.4\% with the baseline, 32.3\% with text-based adaptation, and 42.4\% with audio-based adaptation, showing that text-based adaptation does not hurt performance, while audio-based adaptation significantly improves it.


In brief, our proposed DA method proves effective across different background environments and SNR levels, demonstrating the method's robustness and versatility for broader sound classification tasks. We consistently improve Zero-Shot and TGAP performance. While both text and audio approaches enhance accuracy, audio-based adaptation delivers the most significant gains.

\section{Conclusions and future work}
\label{sec:conclusion}

We show how prototypical audio-text models face challenges in coherently representing background as a concept, which leads to confusion and performance degradation when classifying sounds in real-world settings. To address these challenges, we propose a domain adaptation method that significantly improves classification accuracy in various background environments and SNR levels. Notably, audio-based adaptations outperform text-based approaches, delivering the most substantial gains. Our experiments also underscore the crucial role of modality alignment in zero-shot sound classification, showing that models capable of minimizing the modality gap experience a marked performance boost.
An automatic selection of this parameter $\tau$ is left for future research. 

\bibliographystyle{IEEEtran}
\bibliography{mybib}

\end{document}